\documentclass[12pt]{article} 
\usepackage{epsf}

\newcommand{\ba}{\begin{array}}
\newcommand{\ea}{\end{array}}
\newcommand{\bd}{\begin{displaymath}}
\newcommand{\ed}{\end{displaymath}}
\newcommand{\be}{\begin{equation}}
\newcommand{\ee}{\end{equation}}
\newcommand{\bea}{\begin{eqnarray}}
\newcommand{\eea}{\end{eqnarray}}

\def\r{\rightarrow}
\def\slash {\!\!\!/~}

\def\geta {g_{\eta N\bar N}}
\def\getap {g_{\eta' N\bar N}}

\begin{document}
\setcounter{page}{0}
\thispagestyle{empty}
\setcounter{footnote}{0}
\renewcommand{\thefootnote}{\fnsymbol{footnote}}
\begin{flushright}
{\large MRI-PHY/P981169\\  \tt nucl-th/9811064}
\end{flushright}

\begin{center}
{\Large\bf 
Matter effects on {\boldmath $\eta$} and {\boldmath $\eta'$} mesons
}\\[20mm]
{\em Vivek Kumar Tiwari} \footnote{Permanent address: 
D.D.U. Gorakhpur University, Gorakhpur, India}
and {\em Anirban Kundu} \footnote{Electronic address: 
Anirban.Kundu@cern.ch}
\footnote{Address after 1st December 1998: Department of Physics,
Jadavpur University, Calcutta - 700 032, India}\\[10mm]
{\em Mehta Research Institute,\\
Chhatnag Road, Jhusi, Allahabad - 211 019, India}
\end{center}
\begin{abstract}

We show how the nuclear medium affects the masses of the $\eta$ and $\eta'$ 
mesons.  The change should be easily detectable for dense matter and/or strong 
$\eta(\eta')N\bar N$ coupling. We also find that the $\eta-\eta'$ mixing angle 
is less in magnitude in the nuclear matter than in vacuum. 
\end{abstract}

\vskip 1 true cm
\clearpage
\setcounter{page}{1}
\pagestyle{plain}
\setcounter{footnote}{0}
\renewcommand{\thefootnote}{\arabic{footnote}}

A proper way of understanding the nature of the nuclear force is to study
the interaction between nucleons and low-lying mesons. The nucleon mass 
gets shifted from its vacuum value of $\sim$ 938 MeV due to the radiative
corrections generated by the scalar mesons inside the medium. This correction
has been evaluated in different theoretical frameworks, {\em e.g.}, the
mean-field (MF) model and the relativistic Hartree (RH) model. In turn, 
the meson masses are affected by the nucleon propagators, modified due to
the density-dependent contributions. 

The change of masses for the vector mesons $\rho$ and $\omega$ and
their mixing due to the nuclear medium have been widely investigated 
in the literature\cite{vector}. One of the advantages
for such a treatment for vector mesons is the fairly well-established
(but model-dependent) couplings between the meson and a nucleon-antinucleon
pair \cite{coupling}, 
which leads to a more or less robust prediction. Among the pseudoscalar 
mesons, the coupling of pion to the nucleons can be evaluated from the 
Goldberger-Treiman relation, and is bound to be derivative in nature
since $\pi$ is a Goldstone boson. Such a pseudovector coupling necessarily
generates a small shift for the pion mass. On the contrary,
little is known about the $\eta N\bar N$ and $\eta' N \bar N$ couplings.
A fit to the one-boson exchange potential (OBEP) generates a value of
$\alpha_\eta(=g_{\eta N\bar N}^2/4\pi)$ in the range of 3 to 7 
\cite{kirchbach}, whereas flavour SU(3) relations and chiral 
perturbation theory predict this quantity to be below 1 \cite{scadron}. 
The value of $\alpha_{\eta'}$ is inferred to 
be between $0.25$ and $0.75$ \cite{meissner} from the $pp$ scattering
data.  However, since $\eta$ and $\eta'$ 
are much heavier, these couplings need not be completely pseudovector, and
there can be sizeable pseudoscalar components. 

As we will show,
such couplings introduce significant mass shifts for the $\eta$ and $\eta'$
mesons as well as a mixing between them, analogous to the matter-induced
$\rho-\omega$ mixing discussed in \cite{abhee}. 
This result is valid even if the meson-nucleon coupling is an effective 
one generated through a vertex loop \cite{kirchbach}.
These shifts should be
easily detectable in future hadronic colliders  through the relatively
clean channels $\eta\r 2\gamma$, $\eta\r \ell^+\ell^-\gamma$,
$\eta'\r \rho^0\gamma$, $\eta'\r \omega\gamma$, $\eta'\r 2\gamma$
with a subsequent proper identification of $\rho^0$ and $\omega$.
The branching ratios for the above channels in vacuum are $0.39$, $5
\times 10^{-3} (3\times 10^{-4})$, $0.30$, $0.03$ and $0.021$ respectively
(for the second channel, the first number is for $e^+e^-\gamma$ and the
second one is for $\mu^+\mu^-\gamma$) \cite{pdg}.
We will also show the behaviour of the matter-induced mixing angle 
as a function of nuclear density. The two-photon decay modes of both 
$\eta$ and $\eta'$ are affected by the meson masses as well as the 
mixing angle and can in principle provide a clear testing ground for
such density-dependent effects.

\vskip 1 true cm

The relevant part of the nuclear Lagrangian is \cite{kirchbach}
\be
{\cal L}_{\eta N\bar N} = -i \geta \bar N \gamma_5 N\eta
\ee
where we have neglected the pseudovector coupling, which has a small 
contribution on mass shift and mixing. A similar Lagrangian can be written
for $\eta'$. All the unknown factors, including those arising from an
effective coupling generated by vertex loops, are dumped into
$\geta$.

The nucleon propagator in nuclear matter may be expressed as
\be
G(k) = G_F(k) + G_D(k)
\ee
where
\be
G_F(k)  = (k\slash + M^*)\Big[{1\over k^2-M^{*2}+i\epsilon}\Big]
\ee
and
\be
G_D(k) = (k\slash + M^*)\Big[{i\pi\over E^*(k)}\delta (k_0-E^*(k))
\theta (k_F - |{\bf k}|)\Big]
\ee
with $E^*(|{\bf k}|) = \sqrt{|{\bf k}|^2 + M^{*2} }$, $M^*$ being the
effective mass of the nucleon in the medium.
Since the main contribution to $M^*$
arises from the isospin blind $\sigma$ meson, $M_n^*=M_p^*$.
The term $G_D(k)$ arises from Pauli blocking and describes the modifications
of the propagator at zero temperature by deleting the on-shell propagation
of the nucleon in nuclear matter with momenta below the Fermi momentum $k_F$.
Since we will focus only on symmetric nuclear matter, $k_F^p=k_F^n$; at 
this limit, $\pi^0$ does not mix with either $\eta$ or $\eta'$. 

The radiative correction to the pseudoscalar mass can be written as
\be
-i\Pi (q^2) = \int {d^4k\over (2\pi)^4} (-1) Tr[\gamma_5 iG(k+q) \gamma_5
iG(k)]
\ee
where the negative sign is due to the fermion loop. Writing $G=G_F+G_D$, 
we see that the trace consists of four terms. Among them, the combination 
$G_D(k+q) G_D(k)$ does not contribute in the region in which we are 
interested, and the effect of the free part $G_F(k+q) G_F(k)$, being only
a running of the coupling constant $\geta\r \geta(q^2)$, can be neglected
since $q^2$ itself is quite small in the long-wavelength collective-mode
region \cite{chin}. However, if one wants to see the collective mode over the
entire region of ${\bf q}$ and $q_0$, the free part cannot be neglected;
one needs a suitable subtraction procedure and the implementation of a 
momentum-dependent coupling. One can show that pseudovector couplings
have a small contribution in the long-wavelength limit.

\begin{figure}[htb]
\vspace*{-0.0cm}
\begin{center}
\epsfxsize=9.3cm\epsfysize=7.3cm
\epsfbox{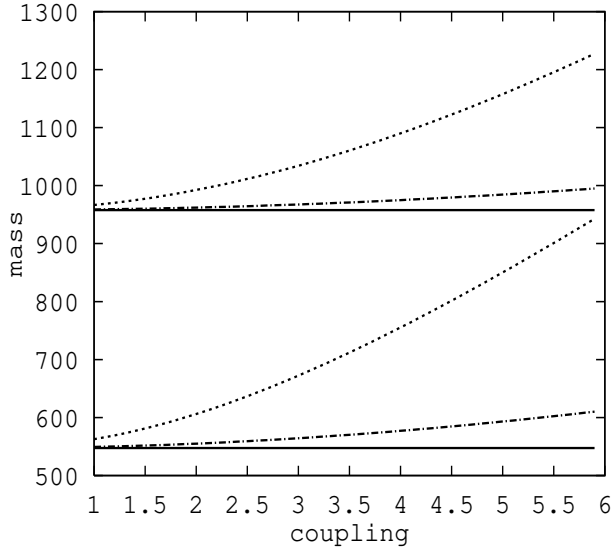}
\vskip -0.3 cm
\end{center}
\caption[fig:fig1]{\em Variation of $\eta$ and $\eta'$ masses with 
coupling $\geta$ ($\getap$). The upper set is for $\eta'$ while the
lower one is for $\eta$. The dot-dashed lines are for $\rho/\rho_0$
(nuclear density ratio) equal to 1 whereas the short-dashed lines
are for $\rho/\rho_0=6$. The horizontal solid lines indicate the 
vacuum masses.
} \label{fig1}
\end{figure}
As we have seen, $\eta$ and $\eta'$ do not mix with $\pi^0$ in the 
symmetric limit. The polarization can be written as a $2\times 2$ matrix
which is of the form
\be
\pmatrix{
1-{\Pi_{\eta\eta}\over q_0^2 - m_\eta^2} &
{\Pi_{\eta\eta'}\over q_0^2 - m_\eta^2}\cr 
{\Pi_{\eta'\eta}\over q_0^2 - m_{\eta'}^2} &
1-{\Pi_{\eta'\eta'}\over q_0^2 - m_{\eta'}^2}
}
\ee
($\Pi_{\eta\eta'}=\Pi_{\eta'\eta}$). The shifted masses are obtained
from the solution of the following equation:
\be
(q_0^2-m_{\eta}^2-\Pi_{\eta\eta}) (q_0^2-m_{\eta'}^2-\Pi_{\eta'\eta'})
-(\Pi_{\eta\eta'})^2 = 0.
\ee
Since the bare masses of $\eta$ and $\eta'$ are quite far apart, we 
neglect their decay widths. The matter-induced mixing angle $\theta_{mix}$
is given by
\be
\theta_{mix} = {1\over 2}\tan^{-1} [{2\Pi_{\eta\eta'}\over (m^{*2}_{\eta}
-m^{*2}_{\eta'})}],
   \label{angle}
\ee
where $m^{*2}_{\eta}=m^2_{\eta}+\Pi_{\eta\eta}$, and same for $\eta'$.

\vskip 1 true cm

\begin{figure}[htb]
\vspace*{-1.5cm}
\hspace*{-2.3cm}
\begin{center}
\epsfxsize=7.55cm\epsfysize=6.6cm
\epsfbox{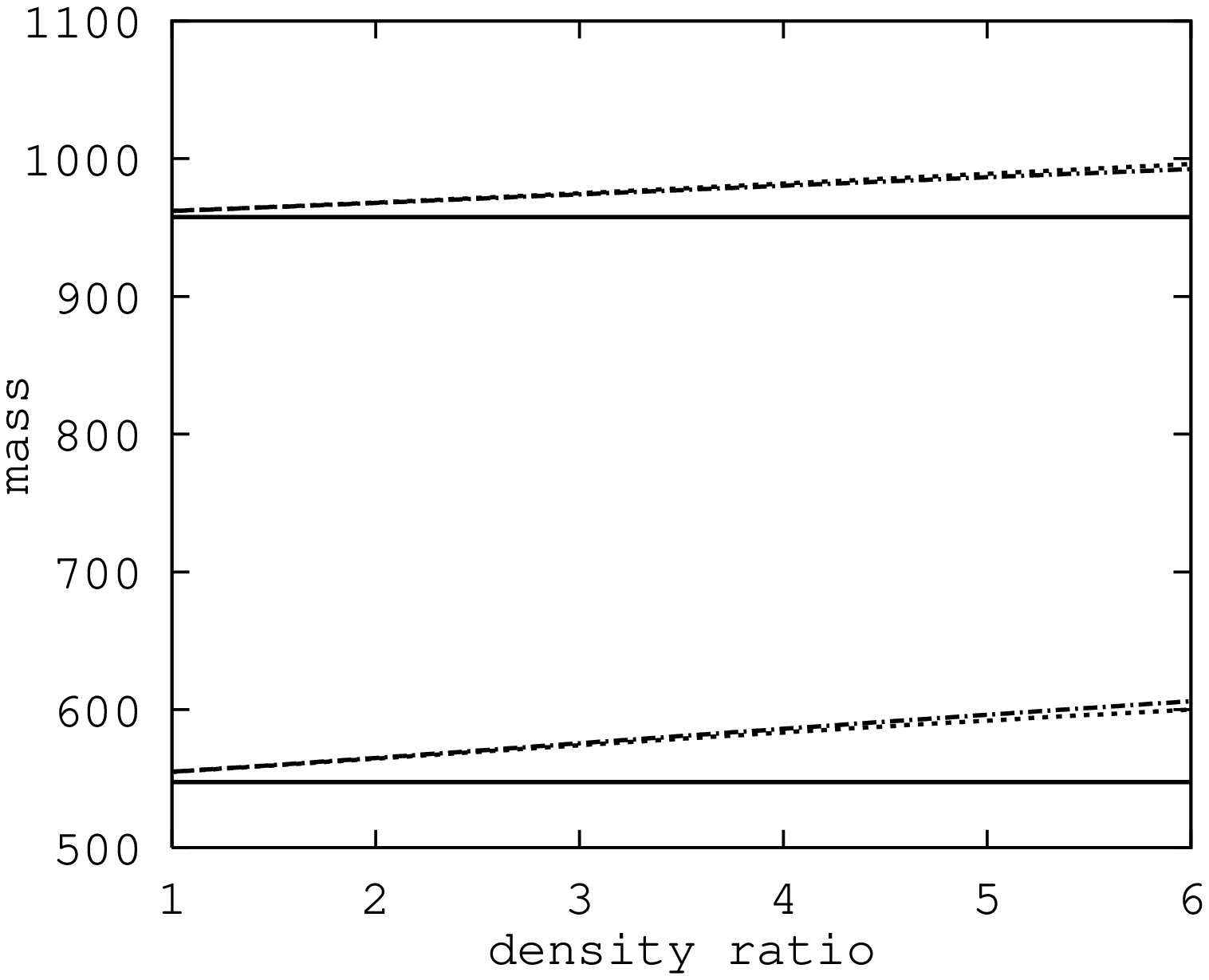}
\hspace*{-.65cm}
\epsfxsize=7.55cm\epsfysize=6.6cm
\epsfbox{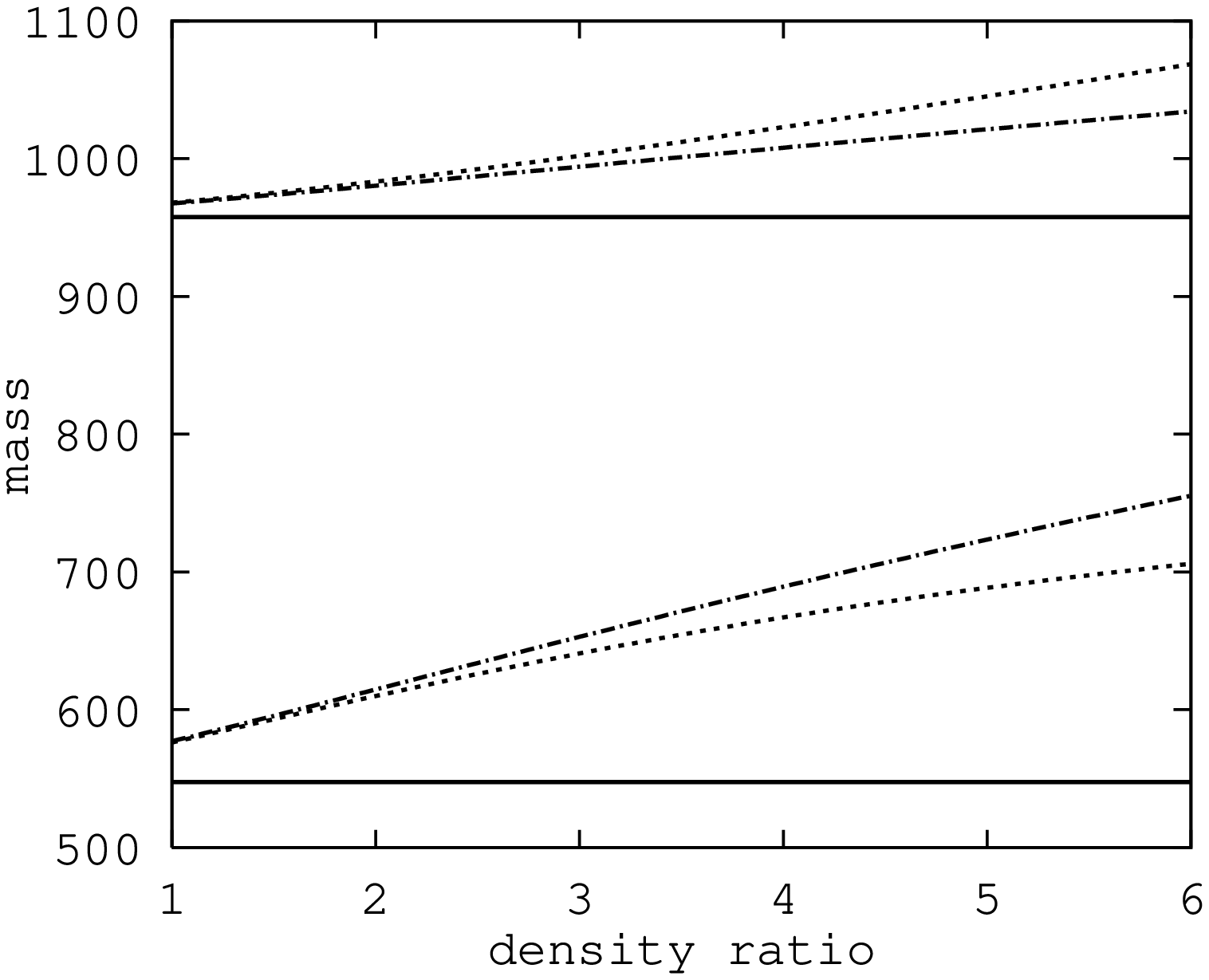}
\vskip -0.6 cm
\end{center}
\caption[fig:fig2]{\em Variation of $\eta$ and $\eta'$ masses with
nuclear density. In both the figures, the upper set is for $\eta'$
and the lower set is for $\eta$. The dot-dashed lines are without 
mixing while the short-dashed lines are with mixing. The horizontal 
solid lines indicate the vacuum masses. In the left-hand figure, 
$\geta=\getap=2$ while in the right-hand figure, $\geta=4$, $\getap=3$.
}
	\label{fig:fig2}

\end{figure}

Now let us discuss our results. In figure 1, we show the variation of
meson masses with couplings $\geta$ or $\getap$ for two different 
densities. We note that the change is significant in the strong coupling 
limit even for ordinary nuclear matter, and for not-so-strong
couplings in dense matter. This is without taking the mixing into account;
as we will see, mixing enhances the $\eta'$ mass and reduces the $\eta$
mass. 
This behaviour is shown clearly in figures 2a and 2b; in figure 2a,
the effect is shown for both $\geta$ and $\getap$ equal to 2, whereas in
figure 2b, $\geta=4$ and $\getap=3$. The range of $\getap$ is more or less 
fixed from ref. \cite{meissner}, but $\geta$ can be larger (however, the 
one-loop Dyson equation approach has no meaning for a very strong 
coupling). 

Assuming both couplings to be positive, we show the plot of the mixing
angle $\theta_{mix}$ with nuclear density in 
figure 3.
\begin{figure}[htb]
\vspace*{-0.0cm}
\begin{center}
\epsfxsize=9.3cm\epsfysize=7.3cm
\epsfbox{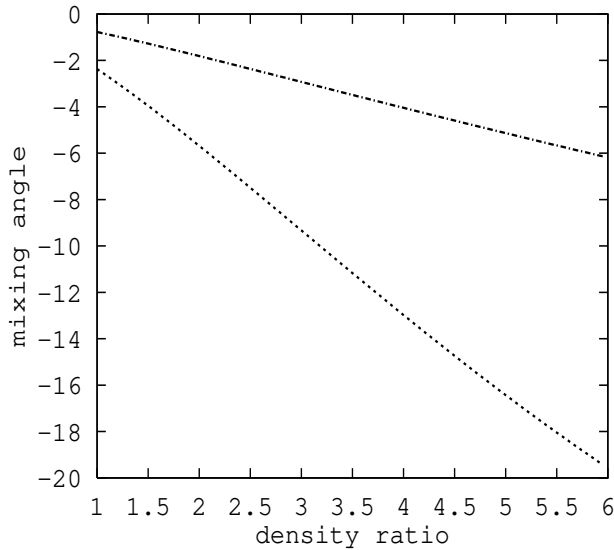}
\vskip -0.3 cm
\end{center}
\caption[fig:fig3]{\em Variation of $\eta - \eta'$ mixing angle
with nuclear density ratio. The dot-dashed line is for $\geta=
\getap=2$ and the short-dashed line is for $\geta=4$, $\getap=3$.
} \label{fig3}
\end{figure}
\begin{figure}[htb]
\vspace*{-0.0cm}
\begin{center}
\epsfxsize=9.3cm\epsfysize=7.3cm
\epsfbox{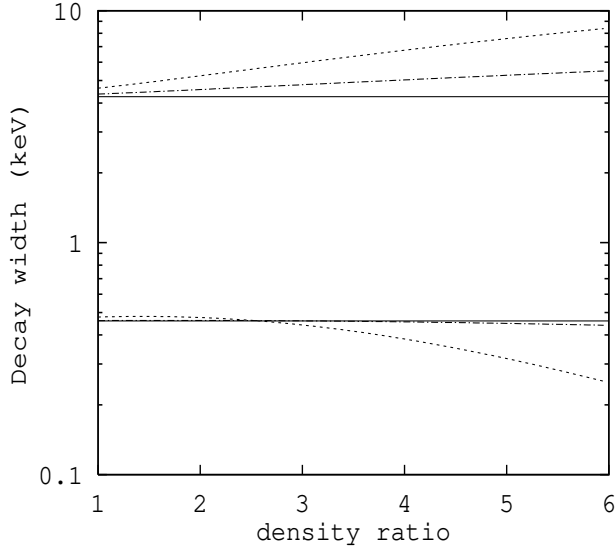}
\vskip -0.3 cm
\end{center}
\caption[fig:fig4]{\em Variation of two-photon decay widths of
$\eta$ and $\eta'$ with nuclear density. The upper set is for $\eta'$
and the lower set is for $\eta$. The dot-dashed lines
are for $\geta=\getap=2$ and the short-dashed lines are for
$\geta=4$, $\getap=3$. The horizontal solid lines denote their
decay widths in vacuum.
} \label{fig4}
\end{figure}
This angle always turns out
to be negative, as the denominator in eq. \ref{angle} contains a negative
quantity, {\em viz.}, $m^{*2}_\eta - m^{*2}_{\eta'}$. Thus, the vacuum 
mixing angle of $\eta$ and $\eta'$, inferred to be $(-21.3\pm 1.5)^\circ$ from
the $\gamma\gamma$ decay mode \cite{ball,ali}, gets reduced in nuclear
matter. In other words, the medium tries to rotate $\eta$ and $\eta'$
back to the gauge basis of $\eta_1$ and $\eta_8$. 
This is a prediction which should be testable in future colliders. 
The two-photon decay width depends on $m_{\eta/\eta'}^3$ and the mixing
angle \cite{ali}. Since both these quantities change in the medium, the
decay width is affected in a nontrivial way. 
We show the two-photon decay width in figure 4. As can be seen, the effect 
is not very prominent for $\eta$ at weak coupling, but for $\eta'$, it should 
be easily detectable. To observe the effect for $\eta$, one needs a strong 
$\geta$ coupling as well as sufficiently dense nuclear matter.

As we have said earlier, 
there are other clean channels to see the matter effects on $\eta$
and $\eta'$, with $\rho$ or $\omega$ being detected in their leptonic
channels.

\vskip 1 true cm

To conclude, we have shown how $\eta$ and $\eta'$ masses change in a 
nuclear matter, where we have assumed the presence of pseudoscalar 
couplings for both these mesons. The change is quite significant and should 
be easily detectable for large nuclear densities and/or strong 
meson-nucleon coupling. The mixing angle is found to be negative compared
to the vacuum mixing angle, but smaller in magnitude than the latter.
The $\eta(\eta')\r\gamma\gamma$ modes can shed more light on this issue.

\end{document}